# Noise-Resilient Detection of Neuronal Spikes by a Hopf-Bifurcation Device

*Jitendra Kumar[1,*], Roberto Fenollosa[1], Gonzalo Rivera-Sierra[1], So-Yeon Kim[1], Adam Armada-Moreira[2], Juan Bisquert[1,*], Michele Giugliano[2,3,4,*]*

[1]Instituto de Tecnología Química (ITQ), Consejo Superior de Investigaciones Científicas-Universitat Politècnica de València, 46022, Valencia, Spain.

[2]Dept. Biomedical, Metabolic and Neural Sciences, Univ. of Modena and Reggio Emilia, 41125 Modena, Italy.

[3]International School for Advanced Studies (SISSA), 34136 Trieste, Italy

[4]National Interuniversity Consortium of Materials Science and Technology (INSTM), 50121 Florence, Italy.

*Corresponding authors E-mail: jkumar2@itq.upv.es, jbisquer@itq.upv.es, mgiugliano@unimore.it

**ABSTRACT**

Detecting weak transient signals in noise is a persistent challenge in sensing, communication, and electrophysiology. Here, we demonstrate a physical weak-signal detector based on a semiconductor negative differential resistance (NDR) device operated near a Hopf bifurcation. A coherent input that persists over the response time of the dynamical system can drive a transition between quiescent and oscillatory states, whereas faster stochastic fluctuations are largely suppressed. This nonlinear transformation converts a weak analog threshold crossing into all-or-none voltage spikes and therefore provides asynchronous signal detection without a reference clock. Using a modulated photovoltaic signal, we detect a weak frequency component of 100 Hz as a demonstration, at an input signal-to-noise amplitude ratio as low as 1/500 (−54 dB) and reliably recover it in the output spectrum. We then apply the same principle to neuronal multisite extracellular recordings. After standard band-pass filtering, the raw microelectrode signal is transformed by the NDR dynamics, enhancing the distinction between neuronal “spikes” and background fluctuations. The resulting detected spike times agree closely with those obtained using a traditional spike-detection pipeline. These results establish bifurcation-engineered NDR dynamics as a compact hardware approach to noise-resilient signal discrimination and event-based analog-to-digital conversion that will be useful for neuroprosthetic devices. In this specific domain, shifting the computational burden of spike discrimination directly into the analog hardware domain would make it possible in the future to circumvent the need for continuous, power-intensive digitization, thereby overcoming the massive data transmission bottlenecks inherent to fully implantable, closed-loop systems.

## INTRODUCTION

Weak signals buried in noise are ubiquitous across nature and technology. Their detection underpins measurements in fields ranging from astronomy and particle physics[1–5] to communications,[6] electrophysiology,[7,8] and machine learning[9–13]. The problem becomes particularly challenging when the signal amplitude approaches or falls below the fluctuations of the measurement system. Under these conditions, instantaneous amplitude alone may be insufficient to distinguish a real event from noise.

Traditional weak-signal detection generally relies on approaches such as amplification, averaging, frequency-selective filtering, matched filtering, synchronous detection, and computational classification. These methods are highly effective when the signal waveform, frequency, or arrival time is known. Brief asynchronous events are more difficult to detect, particularly when signal and noise overlap in amplitude and spectral content. Averaging can erase transient information, while extensive digital processing adds latency, energy consumption, and system complexity.

A nonlinear dynamical system operated near a bifurcation offers a fundamentally different detection principle. Rather than reproducing the input waveform, it can transform a small threshold-crossing input into a large qualitative change of state. An input that acts coherently over the finite response time of the system can drive this transition, whereas shorter stochastic excursions fail to do so. Bifurcation dynamics can therefore combine noise rejection, signal amplification, and event-based analog-to-digital conversion within the same physical device.

Here, we implement this principle with an oscillatory circuit containing a semiconductor thyristor as a negative differential resistance (NDR) element.[14] When an input S drives the system across its bifurcation threshold T, the circuit distinguishes $S < T$ from $S > T$ through the absence or generation of spikes (Figure 1). [14,15] The quiescent and spiking states provide binary outputs, designated 0 and 1, respectively. We demonstrate asynchronous detection of a transient signal at signal-to-noise (SNR) amplitude ratios down to 1/500 (-54 dB), without a reference clock or complex auxiliary circuitry, and show that the device recovers a periodic component that is not visible in the noisy input spectrum. The measurements show that noise rejection and signal amplification occur directly in the physical hardware, avoiding computationally intensive DA/AD hardware or software-based post-processing. Finally, we test the approach on multisite neuronal activity, recorded extracellularly by means of microelectrode arrays (MEAs).

## RESULTS AND DISCUSSION

Threshold discriminators can operate asynchronously and generate digital events, but at very low input signal-to-noise ratios they also respond to noise excursions.[16,17] Lock-in amplifiers can recover substantially weaker periodic signals by narrow-band synchronous detection, but they require a phase or frequency reference and do not natively produce asynchronous event pulses.[18] The present device occupies a different operating regime: it uses its intrinsic dynamics to generate digital spikes without an external clock or phase reference. Table 1 summarizes this qualitative distinction.

Table 1. Qualitative comparison of weak-signal discrimination approaches.

| **Feature** | **Threshold discriminator** | **Lock-in amplifier** | **NDR bifurcation detector** |
|---|---|---|---|
| Sub-unity-SNR operation | No (triggers on noise floor) | Yes (excellent noise rejection) | Demonstrated here at an SNR 1/500 |
| Timing mode | Asynchronous | Requires a frequency/phase reference | Asynchronous; no reference clock required |
| Native output | Digital event | Demodulated analog or digital value | Digital spike or spike- burst |

Because the spike burst envelope follows the coherent modulation of the input, the device is particularly suited to signals whose information is encoded in event timing or frequency. Potential examples include optical event detection,[19,20] clock recovery, and Doppler-based sensing.[21] These examples indicate possible application domains rather than demonstrations performed in the present work.

**Dynamical origin of noise rejection**

The transition between quiescent and oscillatory output is governed by the bifurcation structure described in our previous works.[14,15,22] Near the Hopf boundary, the slowly varying component of the injected current moves the operating point between the stable-equilibrium and stable-limit-cycle regimes, producing quiescent and spiking outputs, respectively. This behaviour is illustrated schematically in Figure 1(c) by a bifurcation diagram in the parameter plane defined by the bifurcation capacitance ($C_{0B}$) and injection current ($I_0$). The red curve, defined by the zero-trace condition ($T_\lambda = 0$) for the Jacobian of the governing system, separates regions of stable ($T_\lambda < 0$) and unstable ($T_\lambda > 0$) equilibria. The horizontal dashed line represents the selected fixed capacitance, and its intersections with the red curve define the bifurcation points. The cyan dot denotes a nominal operating point chosen near the bifurcation boundary, while the double-headed arrow indicates the amplitude of noise-induced fluctuations in the injection current about this point. Noise rejection arises from the finite response time of the state transition. A short noise-induced current fluctuation may cross bifurcation boundary but may not persist long enough to move the system from the stable equilibrium to the stable limit cycle, or vice versa. In this case, the noise only causes a small displacement from the current stable state. The intrinsic relaxation dynamics then drive the system back toward that stable state, effectively suppressing the noise-induced perturbation. It is therefore important to distinguish an instantaneous crossing of the input threshold from the completion of the dynamical transition. The relevant response time is controlled by the internal thyristor dynamics and by the external load capacitance connected in parallel with the device. Together, these parameters define the temporal window over which an input perturbation must act to generate a spike or a burst of spikes. *The circuit consequently acts as a nonlinear temporal discriminator rather than as an instantaneous amplitude comparator.*

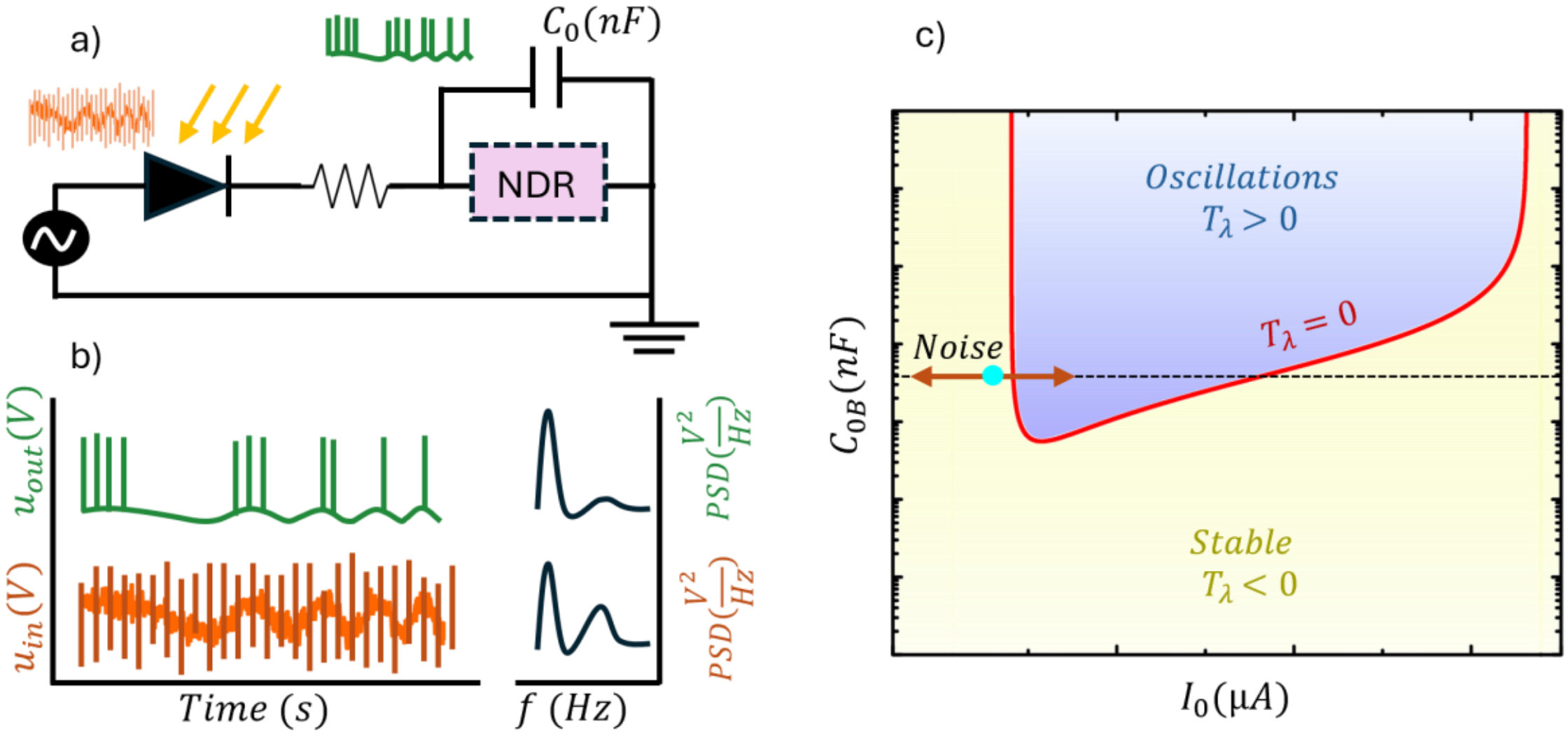

Figure 1. Bifurcation-based weak-signal detection. (a) Schematic of the NDR oscillator circuit. (b) Nonlinear transformation of a coherent threshold-crossing input buried in noise into a readily distinguishable burst of voltage spikes. (c) Bifurcation diagram in the $(I_0, C_{0B})$ parameter plane. The red curve ($T_\lambda = 0$), separates the stable region ($T_\lambda < 0$) from the oscillatory region ($T_\lambda > 0$). The intersections of the dashed line with the red curve define the bifurcation currents for the selected external capacitance. The cyan coloured dot represents the selected operating point near the bifurcation current. The red double-headed arrow indicates noise-induced fluctuations around the injected current.

**Bifurcation-based detection of a weak periodic signal**

Figure 2 demonstrates the detection of a weak time-varying signal. A solar cell illuminated by a modulated light source generates an approximately 20 mV ac voltage component.[15] A waveform generator supplies the dc offset and controlled noise, and the combined signal is applied to the self-oscillator. At an SNR of 2/1, the periodic 100 Hz component is visible in the input, and the output correspondingly alternates clearly between spiking and quiescent states (Figure 2a). The same 20 mV, 100 Hz component is then retained while the noise amplitude is increased systematically (Figure 2(b-d)). Although the 100 Hz modulation becomes indistinguishable by visual inspection of the input, the alternation between output states persists down to an SNR of 1/500. At low mean input voltage, the operating point is sufficiently far from the unstable/oscillatory regime that transient noise excursions do not generate complete spikes. Increasing the dc bias moves the operating point closer to the bifurcation and increases its susceptibility to a sustained input. In the experiments, the small ac photovoltaic component periodically provides this additional drive and initiates a spike burst, whereas shorter incoherent fluctuations are predominantly rejected.

This interpretation accounts for the simultaneous threshold sensitivity and noise tolerance observed in Figure 2. A quantitative treatment of the response time, noise spectrum, and distance from the bifurcation will be required to establish the general detection limits and optimal operating conditions of the device.

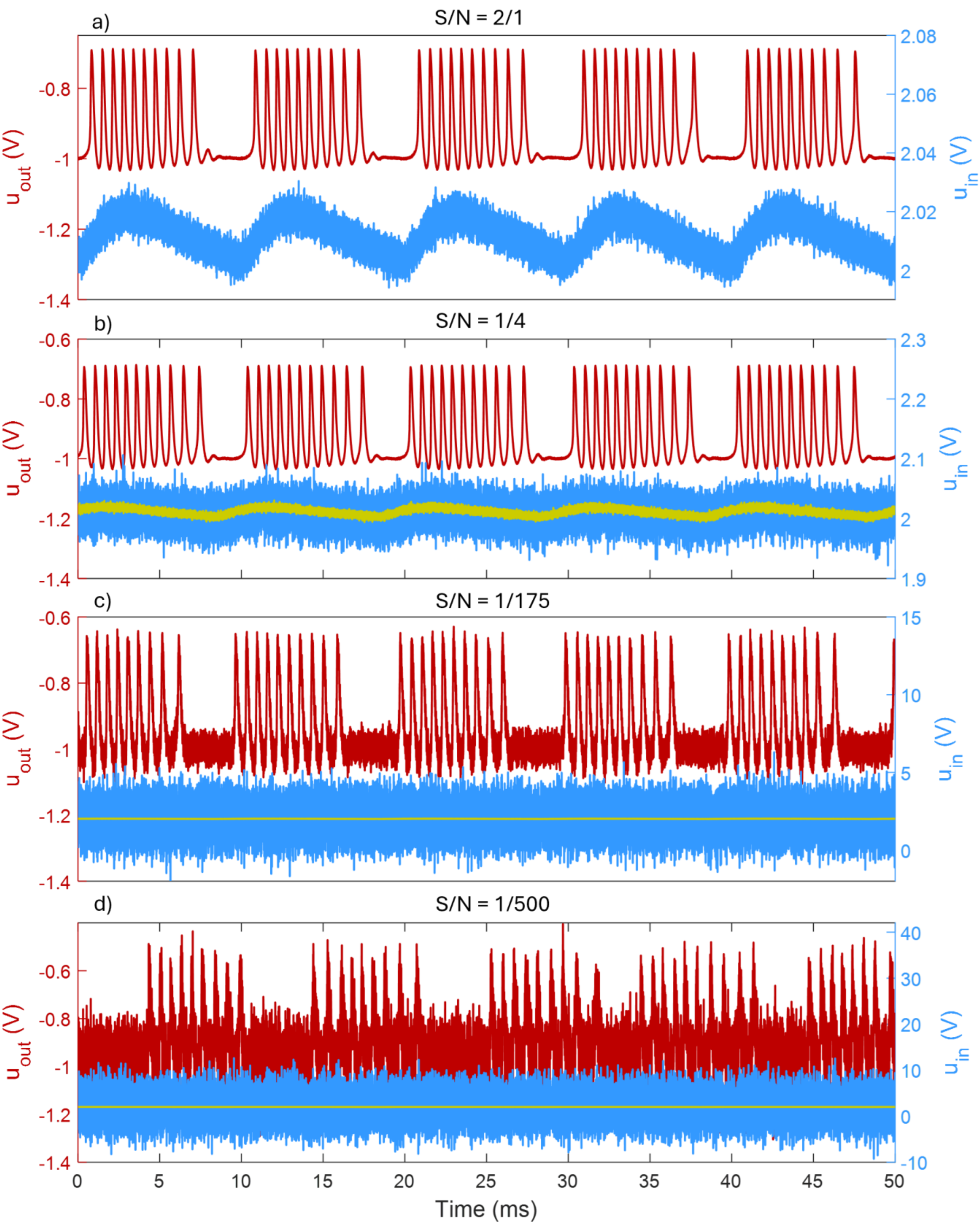

Figure 2. Detection of a weak periodic signal under increasing noise. The output voltage (red) alternates between quiescent and spiking states in response to a 100 Hz, an approximately 20 mV, 100 Hz input component. The input signal-to-noise amplitude ratio is (a) 2, (b) 1/4, (c) 1/175, and (d) 1/500; the coherent input component ($\sim 20\ mV$) is shown in yellow where distinguishable.

**Recovery of the signal frequency below the noise floor**

The frequency-domain analysis provides a more stringent test than visual inspection of the time traces. Figure 3 compares the Fourier spectra of the input and output for two noise levels. When the signal is above the noise floor, both spectra show a peak near 100 Hz. At the higher noise level, the 100 Hz peak is no longer distinguishable in the input spectrum but remains clearly visible in the output. The bifurcation dynamics therefore recover the modulation frequency from an input for which the corresponding spectral component is obscured by noise.

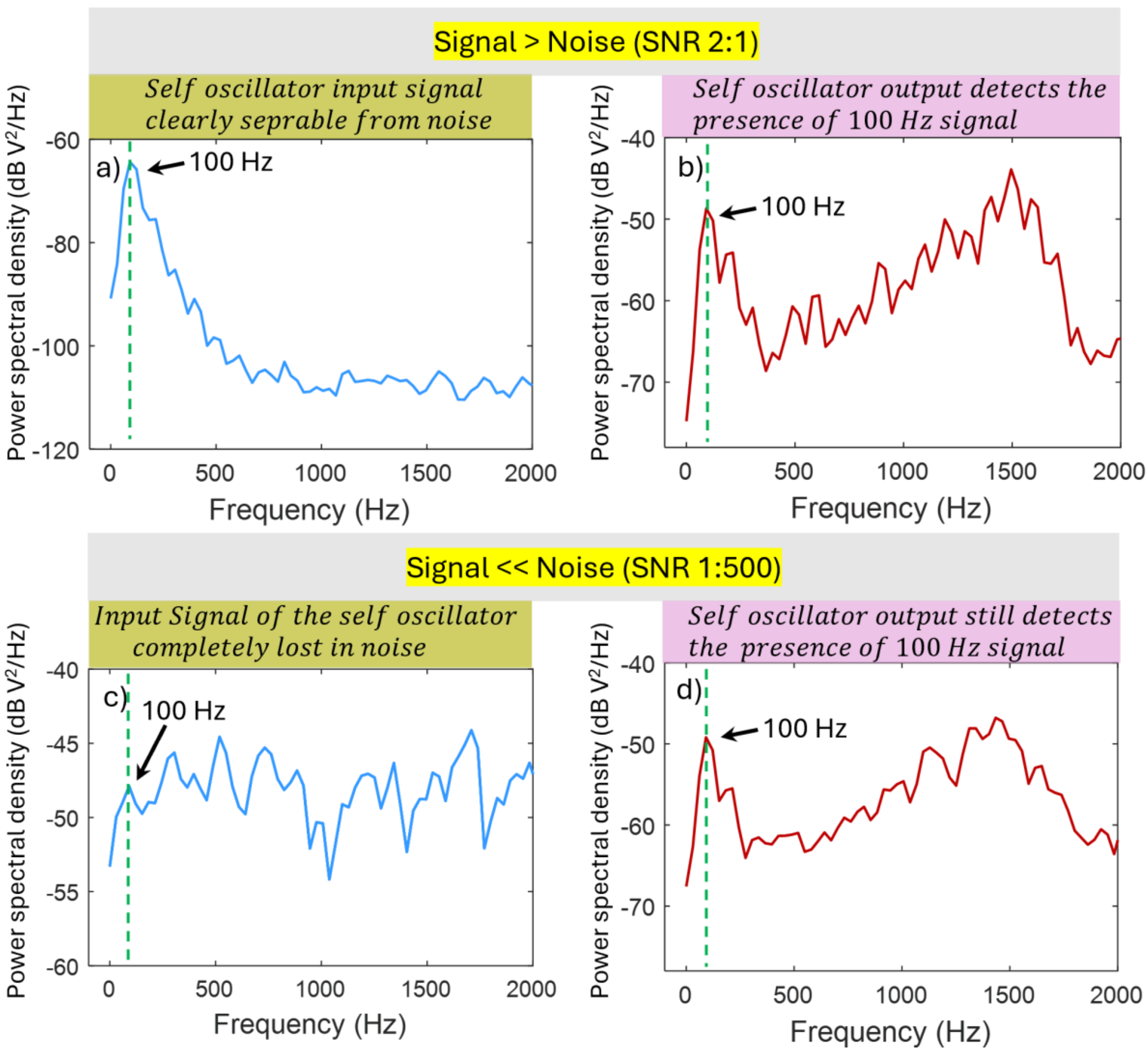

Figure 3. Power spectral densities of the input and NDR-oscillator output. At a signal-to-noise amplitude ratio of 2, the 100 Hz component is visible in both the input (a) and output (b). At a ratio of 1/500, the component is obscured in the input spectrum (c) but remains distinguishable in the output spectrum (d).

**Detection of neuronal spikes in microelectrode-array recordings**

The selective response to temporally coherent inputs suggests extracellular neural recording as a relevant test case. Extracellular action potentials can be comparable in amplitude to the electronic and biological background noise,[8,23–25] and high-density recordings require efficient event extraction. We used recordings from cultured rat primary cortical neurons grown on microelectrode arrays integrated into transparent glass substrates. Brief wide-field optical stimulations were used to evoke collective network responses (Figure 4a).[26]

We first analysed the extracellular recordings with a traditional digital pipeline used here as the reference method. Each channel was zero-phase band-pass filtered from 400 to 3000 Hz to remove slow baseline variations while retaining extracellular spike waveforms.[27] An adaptive threshold (threshold that varied automatically with the background signal) based peak detector identified candidate spikes by distinguishing signal-like peaks from background noise. Thereafter, a polarity-sorting step classified the detected events as positive or negative according to their amplitudes relative to the baseline; in this work, we analyse only the positive spikes. The event times were then displayed as a raster plot to visualize collective neuronal activity. Figure 4(c) shows the first 100 s of the recording; the complete 1800 s raster plot is provided in Supporting Information Figure S1(a).

For a direct comparison, the same band-pass-filtered trace was supplied to the NDR-based dynamical detector. The nonlinear response converted low-amplitude candidate events into pronounced voltage excursions, making subsequent threshold discrimination more robust (Figure 5). Figure 5a and 5c show the band-pass filtered input, and Figure 5d shows the transformed output. Figure 5(a) shows intervals of time at which the noise amplitude exceeds that of genuine spikes, making an adaptive threshold necessary to reliably distinguish spikes from background noise. Red markers denote event times identified by the traditional pipeline (schematically shown in Figure 5b), whereas green-yellow markers denote event times identified

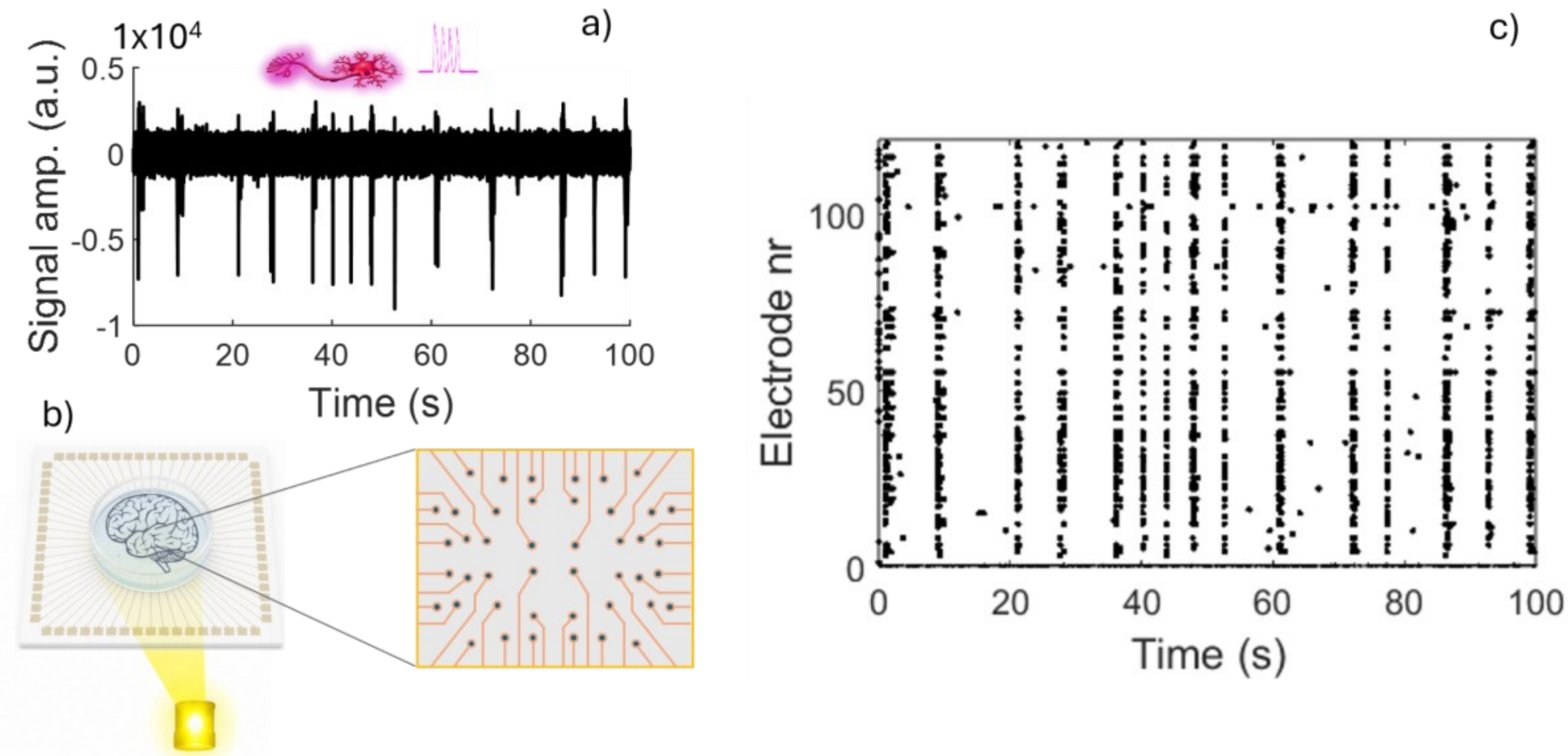

Figure 4. (a) Representative extracellular recording from a single recording channel. (b) Microelectrode array and optical stimulation setup. (c) Event raster plot showing the activity of all 120 recording channels during the first 100 s.

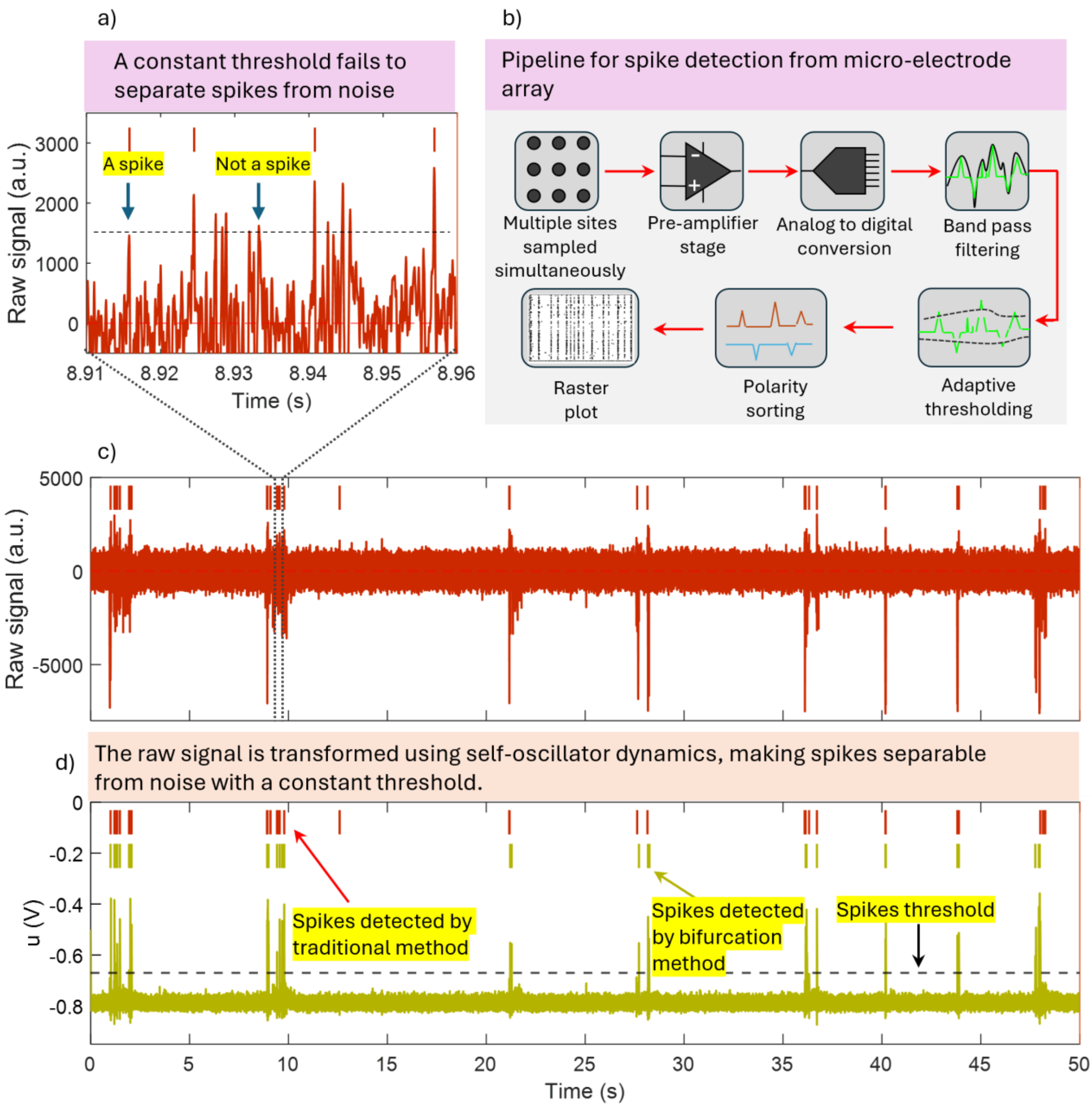

Figure 5. Comparison of traditional and bifurcation-assisted neuronal-event detection. (a) A fixed threshold may not reliably separate all candidate spikes from noise. (b) traditional digital detection pipeline. (c) Band-pass-filtered extracellular trace. (d) Output after nonlinear transformation by the NDR detector. Red and green-yellow markers indicate events detected by the traditional and bifurcation-assisted methods, respectively.

after bifurcation-assisted transformation, which enables the use of a constant threshold.

Both the traditional and bifurcation-assisted nonlinear enhancement methods were applied to the same band-pass-filtered MEA signal to ensure a fair comparison. The bifurcation-based method enhances genuine neuronal spikes while suppressing noise, thereby effectively increasing the SNR and facilitating threshold-based spike detection. The spike times detected by the two methods were plotted together to visually assess their qualitative agreement (Figures 5d and S2c), while cross-correlogram (CCG) analysis provided a quantitative measure of their temporal agreement (Figure S2d). The CCG plot has a peak centred at approximately 250 μs and has a full width at half-maximum (FWHM) of approximately 220 μs (Figure S2d). The narrow FWHM indicates that both methods detect the same neuronal spikes at nearly the same time, while the positive 200 μs shift in the peak centre indicates that the bifurcation-based method consistently detects them slightly later than the other method. This small delay is likely caused by the relaxation dynamics of the oscillator. The signal transformation by oscillator dynamics does not perform spike sorting or identify neuronal units; its function is solely to enhance the SNR for subsequent event-time extraction. Overall, performing asynchronous signal detection without a reference clock presents a substantial advantage for resource-constrained *in vivo* applications. Implementing such a physical weak-signal detector as an on-site preprocessing stage of a neural interface device dramatically reduces the telemetric bandwidth requirements compared to transmitting raw electrophysiological traces.

## CONCLUSIONS

In conclusion, we demonstrated that a semiconductor NDR-based oscillator operated near a Hopf bifurcation can transform a weak coherent input into a binary transition between quiescent and spiking states. Using a modulated photovoltaic input, we detected a 100 Hz component at an SNR 1/500 (−54 dB) and recovered its frequency in the output spectrum without a reference clock. This noise tolerance arises because a given fluctuation must persist over the finite response time of the nonlinear system to complete the transition across the bifurcation boundary and appear as a large voltage excursion; otherwise, relaxation towards the closest attractor reduces the voltage fluctuation amplitude. We further used the device dynamics to transform the band-pass-filtered MEA recordings. The transformation enhanced the amplitude of genuine neuronal events while suppressing noise, thus enabling the reliable extraction of spike times. The spike times detected from the transformed signal closely agreed with those obtained using a traditional detection pipeline. The present results therefore establish a proof of principle of an asynchronous hardware discriminator. Future work should quantify detection probability, false-positive and false-negative rates, latency, energy consumption, and performance on unfiltered multichannel recordings. Ultimately, localized, on-chip spike detection minimizes the data transmission overhead that currently limits the scalability of high-density neuroprosthetics. Such intrinsic noise-resilience and capability for direct on-site signal processing offer an efficient mechanism for developing highly scalable, autonomous brain-machine interfaces.

## EXPERIMENTAL SECTION/METHODS

### Characterization of the coupled solar-cell/thyristor oscillator

A waveform generator, a 470 ohm input resistor, and a solar-cell module were connected in series to inject current into the node containing the thyristor. A 200 nF capacitor was connected in parallel with the thyristor to support and monitor the voltage oscillations. Voltage-time traces were recorded with a PicoScope 4262 oscilloscope (Pico Technology). The output voltage was measured across the parallel thyristor-capacitor branch. The input voltage was characterized separately across the circuit nodes with the thyristor and capacitor removed. Consequently, the reported input-output phase relation is indicative rather than an exact simultaneous phase measurement.

**Cellular electrophysiology and traditional spike-detection**

Primary cortical cell cultures were derived from newborn Wistar rats. All animal experiments were approved by the Italian Ministry of Health, strictly adhering to national and institutional guidelines. Briefly, cells were enzymatically dissociated and seeded onto Polyethyleneimine-coated commercial microelectrode arrays (MEAs) and maintained *ex vivo* in supplemented Modified Essential Medium at 37°C and 5% CO2. Following 21 days in vitro, the spontaneous electrical activity of the developing neuronal networks was monitored and recorded at a sampling frequency of 25 kHz per channel using a multi-channel amplifier setup.

For the data analysis, we employed the pipeline described in (Mahmud et al., 2014, 10.3389/fninf.2014.00026), where raw multiplexed recordings were processed in MATLAB, included a fourth-order zero-phase band-pass filtering between 400 and 3000 Hz, median-based peak detection for extracting multi-unit activity time-stamps, and elementary spike sorting based on event polarity alone.

**Conflict of Interest**

The authors declare no competing financial interest.

**ASSOCIATED CONTENT**

**Supporting Information**

Current-voltage characteristics of the thyristor and solar-cell module; variation of intra-burst frequency; and dependence of the spike rate on input frequency, amplitude, and dc offset.

**Data Availability Statement**

The data presented here can be accessed at https://doi.org/10.5281/zenodo.22247773 (Zenodo) under the license CC BY 4.0 (Creative Commons Attribution 4.0 International).

**AUTHOR INFORMATION**

**Corresponding Authors**

Dr. Jitendra Kumar − Instituto de Tecnología Química (ITQ), Consejo Superior de Investigaciones Científicas-Universitat Politècnica de València, 46022, València, Spain. orcid.org/0000-0002-5530-201X; Email: jkumar2@itq.upv.es

Prof. Juan Bisquert − Instituto de Tecnología Química (ITQ), Consejo Superior de Investigaciones Científicas-Universitat Politècnica de València, 46022, València, Spain. orcid.org/0000-0003-4987-4887; Email: jbisquer@itq.upv.es

Prof. Michele Giugliano − Dept. Biomedical, Metabolic and Neural Sciences, Univ. of Modena

and Reggio Emilia, 41125 Modena, Italy; International School for Advanced Studies (SISSA), 34136 Trieste, Italy; National Interuniversity Consortium of Materials Science and Technology (INSTM), 50121 Florence, Italy. orcid.org/ 0000-0003-2626-594X; Email: mgiugliano@unimore.it

**ACKNOWLEDGMENTS**

This work was funded by the European Research Council (ERC) via Horizon Europe Advanced Grant, grant agreement nº 101097688 ("PeroSpiker"). MG acknowledges support from the European Innovation Council (EIC) via Horizon Europe Pathfinder programme, grant agreement nº 101070908 ("CrossBrain"). The authors gratefully acknowledge additional institutional support from the Spanish Ministry of Science and Innovation (grant CEX2021-001230-S, funded by MCIN/AEI/10.13039/501100011033) and the University of Modena and Reggio Emilia.

# Noise-Resilient Detection of Neuronal Spikes by a Hopf-Bifurcation Device

*Jitendra Kumar[1,*], Roberto Fenollosa[1], Gonzalo Rivera-Sierra[1], So-Yeon Kim[1], Adam Armada-Moreira[2], Juan Bisquert[1,*], Michele Giugliano[2,3,4,*]*

[1]Instituto de Tecnología Química (ITQ), Consejo Superior de Investigaciones Científicas-Universitat Politècnica de València, 46022, Valencia, Spain.

[2]Dept. Biomedical, Metabolic and Neural Sciences, Univ. of Modena and Reggio Emilia, 41125 Modena, Italy.

[3]International School for Advanced Studies (SISSA), 34136 Trieste, Italy

[4]National Interuniversity Consortium of Materials Science and Technology (INSTM), 50121 Florence, Italy.

*Corresponding authors E-mail: jkumar2@itq.upv.es, jbisquer@itq.upv.es, mgiugliano@unimore.it

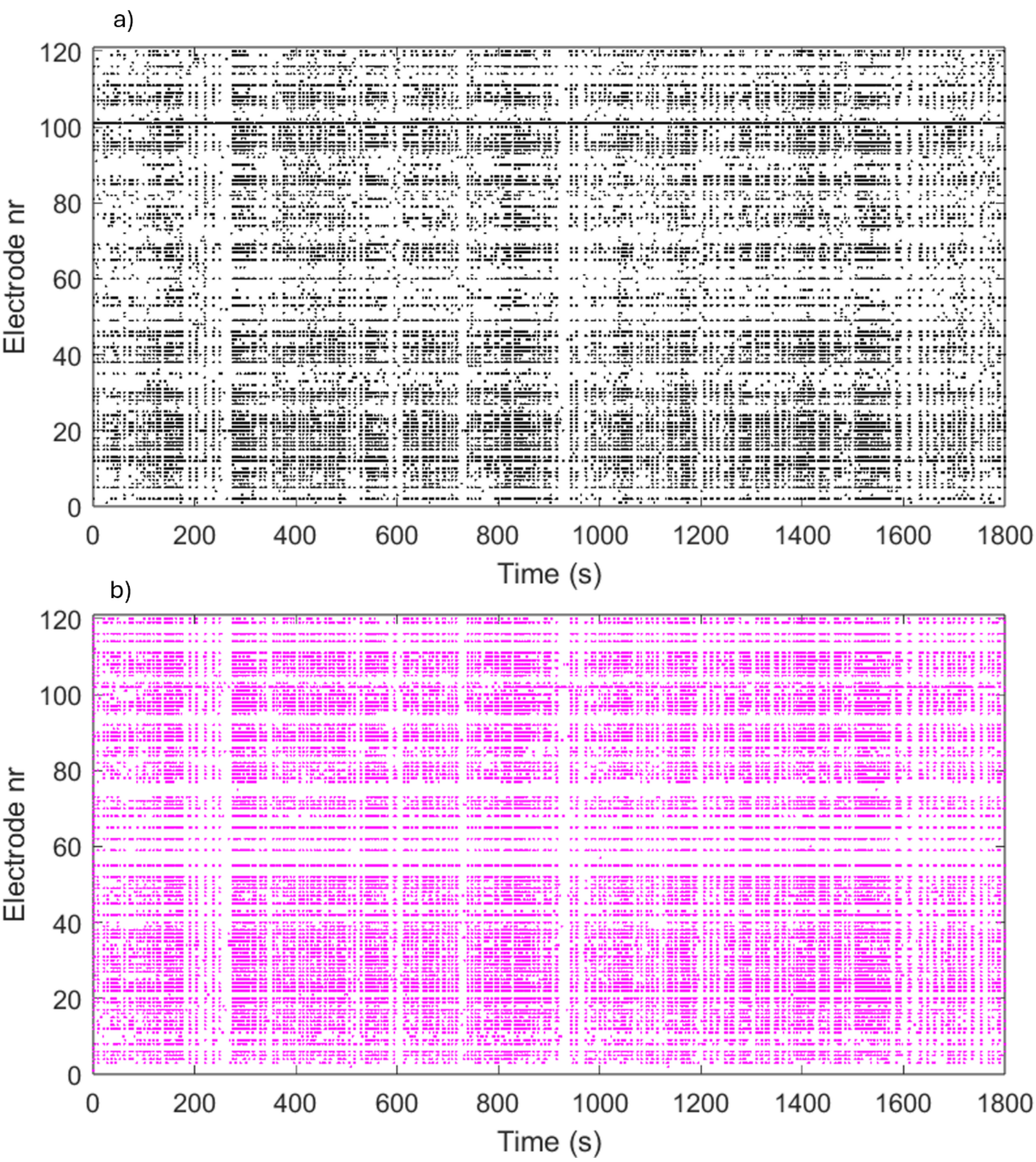

Figure S1. Raster plot showing activity across all 120 electrodes over a 30-minute recording period. (a) Raster plot of spike-times detected by traditional method, (b) spike-times detected by our method.

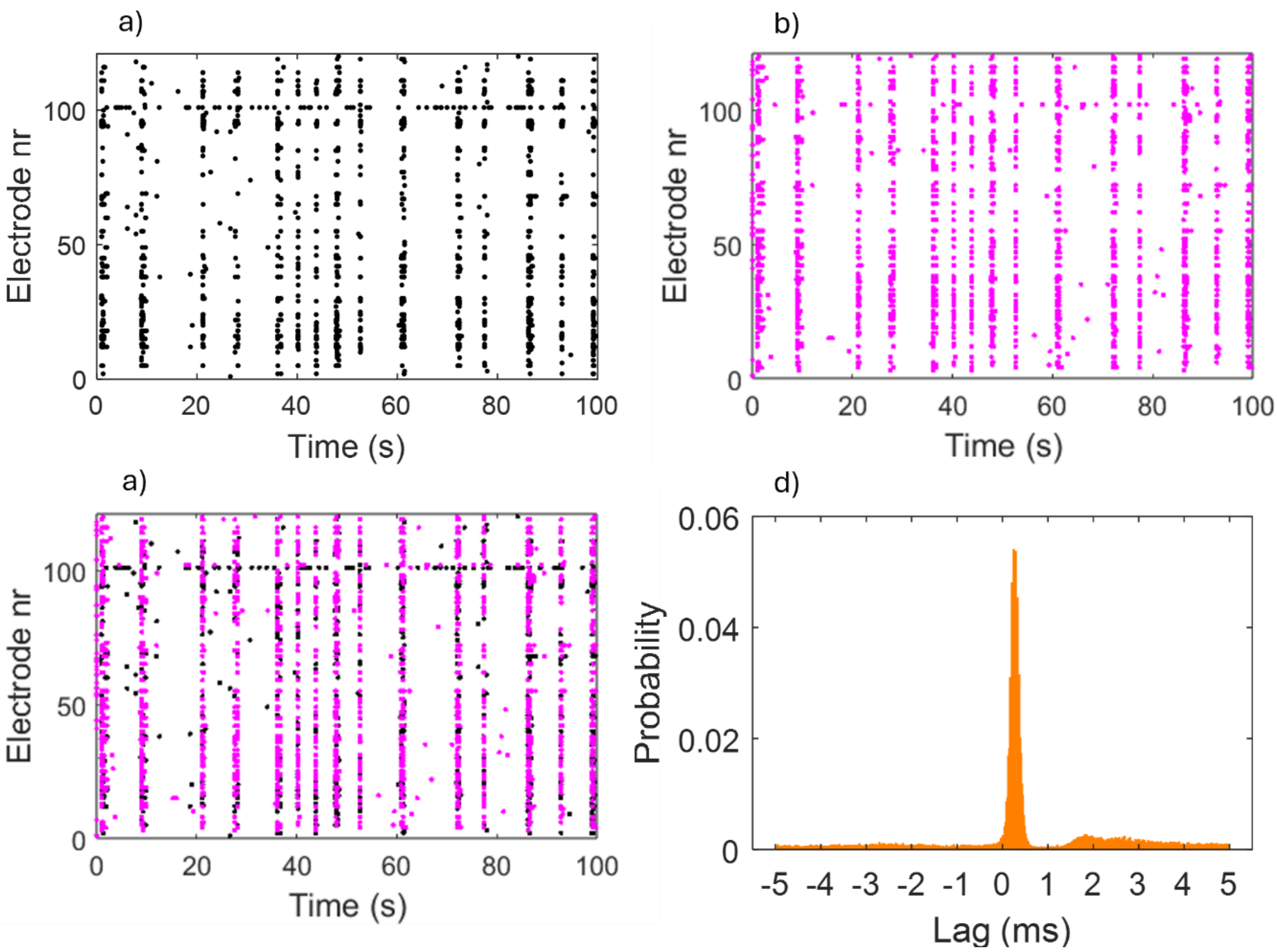

Figure S2. Comparison of spike times detected using the traditional method and our method. (a) Spike times detected by the traditional method, (b) spike times detected by our method, (c) overlap of spike times detected by the two methods, (d) cross-correlogram (CCG) showing a narrow peak around 0.25 ms, indicating that our method detects spikes at nearly the same times as the traditional method.

**MATHEMATICAL MODEL OF OSCILLATOR DYNAMICS**

The dynamics of the current-driven thyristor oscillator are described by the following nonlinear system:

$$\frac{du}{dt} = \frac{1}{C_0}[I_{in}(u,t) - g(x)u] \tag{1}$$

$$\frac{dx}{dt} = \frac{1}{\tau_k}[g(x)u - x] \tag{2}$$

where $C_0$ is capacitance, $\tau_k$ is current relaxation time.

$I_{in}(u,t)$ is the input signal that needs to be nonlinearly transformed, in microelectrode array (MEA) application. The electrode voltage needs to be mapped into the acceptable range of current required by the thyristor system.

The nonlinear conductance of the thyristor is defined as

$$g(x) = \frac{x}{U_{eq}(x)}$$

and the equilibrium current – voltage current–voltage characteristic is

$$U_{eq}(x) = R_v x + V_0 ln\left|\frac{d_0^2 a_0 + d_0 a_1 x + a_2 x^2}{d_0 b_0 + b_1 x + b_2 x^2}\right| + V_1 \tag{3}$$

where, $R_v$, $V_0$, $V_1, d_0, a_1, a_2, b_0, b_1, b_2$ are thyristor fitting parameters.

The Jacobian of the dynamical system is written as,

$$J = \begin{bmatrix} \left[\frac{\partial I_{in}(u,t)}{\partial u} - g(x)\right]/C_0 & -(u/C_0)\cdot g'(x) \\ g(x)/\tau_k & (u \cdot g'(x) - 1)/\tau_k \end{bmatrix}$$